\documentclass{ws-procs11x85}

\makeindex
\begin{document}

\title{NEUTRINO PHYSICS: 
OPEN THEORETICAL QUESTIONS}

\author{A. Y. SMIRNOV}

\address{International Centre for Theoretical Physics, Strada Costiera 11,
31014 Trieste, Italy \\and\\
Institute for Nuclear Research,  Russian Academy of Sciences,Moscow, Russia
 \\E-mail: smirnov@ictp.trieste.it}


\twocolumn[\maketitle\abstract{We know that   
neutrino mass and mixing provide a window to physics beyond the Standard Model. 
Now this window is open, at least partly. And the questions are: what do we see, 
which kind of new physics, and how far ``beyond"? 
I summarize the present knowledge of neutrino mass and mixing,  
and then formulate the main open questions. Following the bottom-up approach,    
properties of the neutrino mass matrix are considered. Then 
different possible ways to uncover the underlying physics are discussed.   
Some  results along the line of: see-saw,  GUT and SUSY GUT are reviewed.}]

\baselineskip=13.07pt
\section{Introduction}

This review~\footnote{Talk given at {\it the XXI International 
Symposium on Lepton and Photon Interactions at 
High Energies, ``Lepton Photon 2003"}, August 11-16, 2003 - Fermilab, Batavia, IL USA.}
 is devoted to neutrino masses and mixing.  It covers   
experimental results, their interpretation and implications. 
It is in this area that enormous progress has been achieved 
during the last few years. 

The field develops fast,  and already after the Symposium a number of  important 
results have been published
including the SNO salt phase data, new analysis of the 
Heidelberg-Moscow experimental results, {\it etc.}. 

In Sec. 2 the main achievements in reconstruction of the neutrino mass and mixing 
spectrum are summarized. The open theoretical questions are formulated in Sec. 3. 
In Sec. 4, following the bottom-up approach, the neutrino mass matrix is reconstructed 
and its properties  are studied. In Sec. 5 the ways we may go in answering 
the open questions  are outlined.

\section{What Have We Learned?}

\subsection{\it \bf  Solar Neutrinos}
\label{subsec:sol}

\begin{figure}[h!]
\begin{center}
\hspace{-0.1cm} \epsfxsize7cm\epsffile{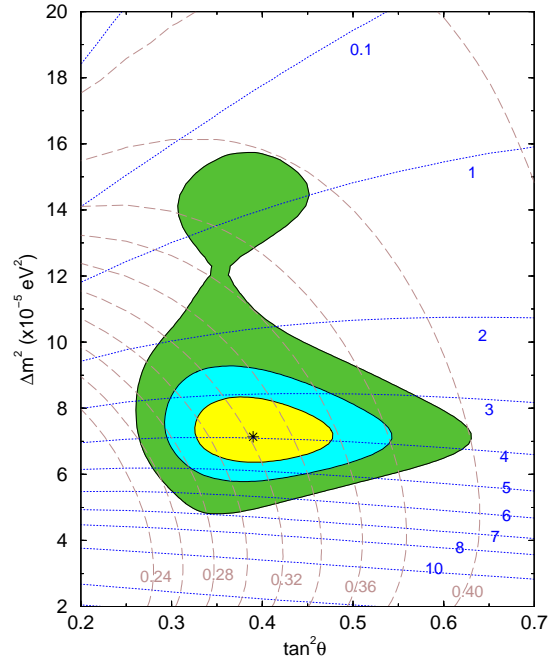}
\vspace{-3mm}
\caption{The allowed regions of oscillation parameters from the combined fit of the solar 
neutrino data and the KamLAND spectrum at $1\sigma$, $2\sigma$, $3\sigma$ CL.$^{10}$ 
Shown are also the contours of constant CC/NC 
ratio (dotted lines) and the Day-Night asymmetry (dashed lines) 
at SNO (numbers on the curves in \%).}
\label{allowed}
\end{center}
\end{figure}

The latest SNO  salt phase results\cite{salt}
have  further confirmed the correctness of the Standard Solar Model (SSM) neutrino 
fluxes\cite{ssm} and the realization of the MSW large mixing (LMA) conversion 
mechanism\cite{msw} 
inside the Sun.\cite{balan}$^{-}$\cite{pedro3} 
In Fig.~\ref{allowed} we show the allowed region of  the oscillation 
parameters  $\tan^2 \theta_{12}$ and $\Delta m^2_{12}$ from the 
$2\nu$ combined analysis of the solar neutrino and KamLAND\cite{KL} results. 
The best-fit values of the parameters are    
\begin{equation}
\Delta m^2_{12} = 7.1 \times 10^{-5} {\rm eV}^2, ~~ 
\tan^2 \theta_{12} = 0.4. 
\label{eq:bf}
\end{equation}

\vspace{-5mm}

Combined fit of the solar, KamLAND\cite{KL} and CHOOZ\cite{chooz} results favors nearly  
zero 1-3 mixing: $\sin^2 \theta_{13} \sim 0$\rlap{.}\,\cite{choubey,pedro3} 
Basically the data have selected the l-LMA  
region with $\Delta m^2_{12} < 10^{-4}$ 
${\rm eV}^2$ (the h-LMA region is accepted now at $3\sigma$ only),  
and strongly disfavored maximal 1-2 mixing. The upper bound is   
\begin{equation}
\tan^2 \theta_{12} <  0.64 ~~~~~(3 \sigma). 
\label{eq:up12}
\end{equation}
As a result of these improvements, the 
physics of the conversion is now even determined quantitatively\rlap{.}\,\cite{fogli,pedro3} 
In particular, recent results show  relevance of the notion of resonance,  they fix  
the relative strength of the effects of the adiabatic conversion 
and the oscillations as function of the neutrino energy\rlap{.}\,\cite{pedro3} 

In Fig.~\ref{allowed} we show also the contours of constant CC/NC ratio and 
Day-Night asymmetry of the CC-events at SNO. They allow one to evaluate an impact of future SNO measurements. 
The KamLAND operation will allow one to eventually determine 
$\Delta m^2_{12}$ with about 10\% accuracy. 
  
Are there any  data which indicate deviation from the LMA picture?  
In this connection  we consider  two generic features  
of the LMA-MSW solution:  
\begin{itemize}

\item 
the predicted Ar-production rate,   
$Q_{Ar} = 2.96 \pm 0.25$ SNU,   is 
about $2\sigma$ higher than the Homestake\cite{Cl} result; and

\item
the upturn of the spectrum at low energies, 
that is, the increase of the ratio $N^{obs}/N^{SSM}$  with decrease of energy,  
is expected which can be as large as 10\,--\,15\%. However, the latest 
SNO as well as  the previous SNO and SuperKamiokande\cite{SKs} 
spectral data do not show the  upturn, being  in agreement 
with the absence of distortion. 

\end{itemize}


\begin{figure}[h!]
\begin{center}
\hspace{-0.1cm} \epsfxsize7.5cm\epsffile{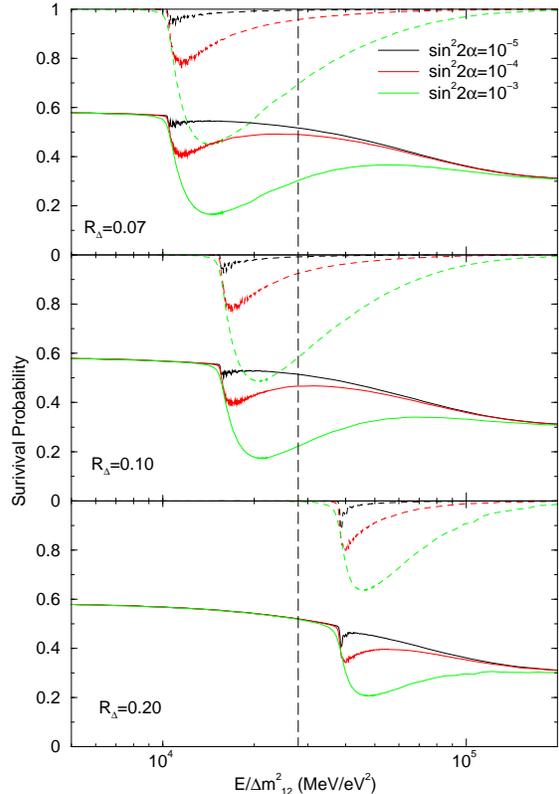}
\caption{The survival probabilities of the electron neutrino (solid line) and  the active 
neutrinos (dashed line) for different values of the sterile-active mixing.$^{15}$ 
The panels correspond to three different values of 
$R_{\Delta} \equiv \Delta m_{01}^2/\Delta m_{21}^2$. 
Vertical dashed line indicates the position of the 1-2 resonance. 
} 
\label{dip}
\end{center}
\end{figure}

Both  problems can be resolved simultaneously, if a  light sterile neutrino exists with 
very small active-sterile mixing:\,\cite{ster} 
\begin{equation}
\frac{\Delta m_{01}^2}{\Delta m_{21}^2} =  0.05 - 0.2, ~~~ \sin^2 2\alpha = 
10^{-5} - 10^{-3}. 
\label{eq:it}
\end{equation}
Such a mixing produces a dip in the survival probability (Fig.~\ref{dip}) which suppresses   
both the Ar-production rate and  the upturn of  spectrum. 
The best  description of the data would correspond 
to the dip at relatively high energies 
(panel for $R_{\Delta} = 0.10$) when the CNO- and pep-neutrino fluxes 
and the low energy part of the  boron neutrino spectrum are suppressed. 

Such a possibility  can be tested in the future low energy neutrino experiments: 
BOREXINO\rlap{,}\,\cite{bor} KamLAND, MOON, {\it etc.}\rlap{,}\,\cite{moon} as well as in further 
measurements of the spectrum by SNO and SK.

\subsection{\bf Atmospheric Neutrinos}

\begin{figure}[h!]
\begin{center}
\hspace{-0.1cm} \epsfxsize8cm\epsffile{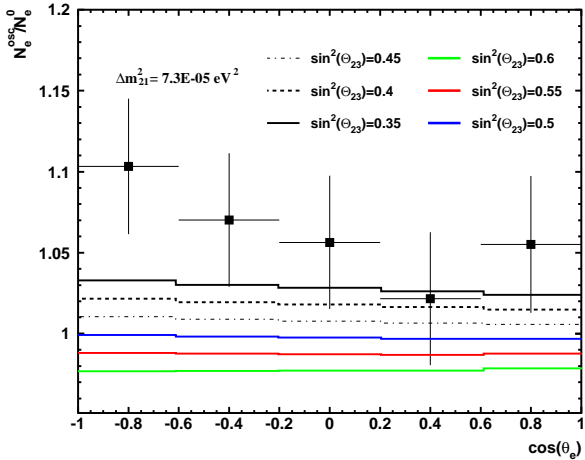}
\vskip -6cm
\caption{The ratio of numbers of the $e$-like events with and without oscillations 
as function of the zenith angle of the electron for  different values of $\sin^2 \theta_{23}$.$^{20}$ 
Other parameters are $\sin^2 2\theta_{12} = 0.82$, $\sin \theta_{13} = 0$ and 
$\Delta m^2_{12} = 7.3 \times 10^{-5}$ eV$^2$.  Also  shown are the 
SuperKamiokande experimental points. }
\label{atm}
\end{center}
\end{figure}

A recent refined analysis of the SuperKamiokande data in terms 
of $\nu_{\mu} - \nu_{\tau}$ oscillations gives\cite{atm} at 90 \% C.L. 
\begin{equation}
\Delta m^2_{13} = (1.3 - 3.0) \times 10^{-3} {\rm eV}^2, ~~
\sin^2 2\theta_{23} >  0.9 ~
\label{eq:atm}
\end{equation}
with the best fit at $\Delta m^2_{12} = 2.0 \times 10^{-3}$ eV$^2$ and $\sin^2 2\theta_{12} = 1.0$. 
Combined analysis of the CHOOZ and the atmospheric neutrino data puts the upper bound  
on the  1-3 mixing\cite{atmfo} 
\begin{equation}
\sin^2 \theta_{13} <   0.067 ~~~~(3\sigma). 
\label{eq:atm}
\end{equation}
The open question is whether oscillations of the 
atmospheric $\nu_e$ exist? There are two possible sources of these oscillations: 
(i) non-zero 1-3 mixing and ``atmospheric" $\Delta m^2_{13}$, and 
(ii) solar oscillation parameters in Eq.~(\ref{eq:bf}).  
Also their interference  should exist\rlap{.}\,\cite{PS-L} After  confirmation of the 
LMA-MSW solution we can definitely say that oscillations driven by the LMA parameters 
(the LMA oscillations) should show up at some level. 
Relative modification of the $\nu_e$ flux due to the LMA oscillations can be 
written as\cite{PS-L} 
\begin{equation}
\frac{F_e}{F_e^0} - 1 = P_2 (r \cos^2 \theta_{23} - 1), 
\label{eq:atm-e}
\end{equation}
where $P_2(\Delta m^2_{12}, \theta_{12})$ is the $2\nu$ transition probability and 
$r \equiv {F_{\mu}^0}/{F_e^0}$ is the ratio of the original $\nu_{\mu}$ and $\nu_e$ fluxes. 
In the sub-GeV region, where $P_2$ can be of the order 1, the ratio equals $r \approx 2$,  so that 
the oscillation effect is proportional to the deviation of the 2-3 mixing from the
maximal value: $D_{23} \equiv 1/2 - \sin^2 \theta_{23}$.    
In  Fig.~\ref{atm}  
we show the ratio of numbers of the $e$-like events with and without oscillations
as function of the zenith angle of the electron. 
For the allowed range  of $\sin^2 \theta_{23}$ and the present best-fit value of 
$\Delta m^2_{12}$ the excess can be as large as 5 - 6\%. 
The excess increases with decreasing energy.  
\vskip 0.7cm

\vskip -0.5cm

Future  searches for the excess can be  used to restrict or measure $D_{23}$. 
In fact, the latest analysis, 
(without  renormalization of the original fluxes) shows some excess of the $e$-like events at 
low energies and the absence of excess in the multi-GeV sample, thus  giving a
hint of non-zero $D_{23}$.  
Establishing this deviation has important consequences for understanding the origins 
of neutrino masses and mixing. 

Non-zero 1-3 mixing generates the interference effect  
between the LMA oscillations amplitudes\rlap{.}\,\cite{PS-L}  
The interference contribution  does not contain the ``screening" factor, 
in Eq.~(\ref{eq:atm-e}), and can reach 2\,--\,4\% for the allowed values of $\sin \theta_{13}$.  
This produces an uncertainty in the determination of $D_{23}$.  
So,  $D_{23}$ can be measured if either a large excess is found  or/and a stronger bound on 
the 1-3 mixing is established.   

\subsection{\bf Mass Spectrum and Mixing}

\begin{figure}[h!]
\begin{center}
\hspace{-0.1cm} \epsfxsize8cm\epsffile{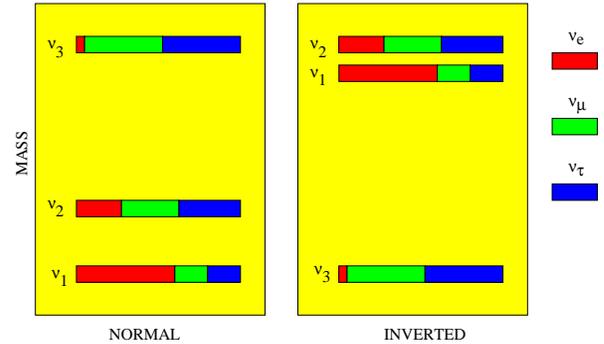}
\caption{Neutrino mass and flavor spectra for the normal (left) and inverted (right) 
mass hierarchies. The distribution of flavors (colored parts of boxes) in the mass eigenstates 
corresponds to the best-fit values of mixing parameters  and  $\sin^2 \theta_{13} = 0.05$. 
}
\label{sp}
\end{center}
\end{figure}

Information obtained 
from the oscillation experiments 
allows us  to make significant progress in the reconstruction of the neutrino mass and flavor  
spectrum  (Fig.~\ref{sp}). 

The unknowns are: 

(i) admixture of $\nu_e$ in $\nu_3$:  $U_{e3}$; 

(ii) type of mass spectrum:
hierarchical; non-hierarchical with certain ordering; degenerate, 
which is related to the value of the absolute mass scale, $m_1$; and

(iii) type of mass hierarchy (ordering): normal, inverted. 

Using a global fit of the oscillation data one can find  intervals for the 
elements of the PMNS mixing matrix $||U_{\alpha i}||$: 
\begin{equation}
\left(
\begin{tabular}{lll}
0.79 - 0.86 & 0.50 - 0.61 & 0.0 - 0.16\\
0.24 - 0.52 & 0.44 - 0.69 & 0.63 - 0.79\\
0.26 - 0.52  & 0.47 - 0.71 & 0.60 - 0.77\\
\end{tabular}
\right), 
\label{eq:it}
\end{equation}
where columns  correspond to the flavor index and rows to the mass index\rlap{.}\,\cite{concha} 

\begin{figure}[h!]
\begin{center}
\hspace{-0.1cm} \epsfxsize7.5cm\epsffile{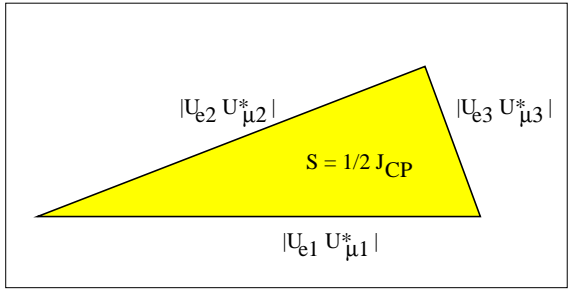}
\caption{Possible leptonic unitarity triangle. We take the best-fit values of  
$\theta_{12}$,  and $\theta_{23}$ and $\sin \theta_{13} = 0.16$.}
\label{tri}
\end{center}
\end{figure}

Now we are in a position to construct the leptonic unitarity triangle, although 
the finite size of one angle is  still unknown. For practical 
reason (no intensive $\nu_{\tau}$ beams) we consider the triangle which employs the 
$e$- and $\mu$- rows of the mixing matrix  (Fig.~\ref{tri}).  
The triangle is not degenerate in spite of the strong bound on the 1-3 mixing. 

Is it possible to  reconstruct the triangle using results from  future 
experiments? Can we use the triangle to determine the CP-violation phase, $\delta$? 
The area of the triangle is related to  the Jarlskog invariant  
$J_{CP} \equiv Im[{U_{e1} U_{\mu2} U_{e2}^* U_{\mu1}^*}]$:
$S = J_{CP}/2$. Reconstruction of the triangle is complementary 
to  measurements of the neutrino-antineutrino asymmetries in oscillations. 
Interestingly, the main problem here is the coherence: the same coherence 
which leads to the oscillations. For the triangle method we need to 
study interactions of the mass eigenstates, whereas in practice we deal  with
flavor (coherent) states. So, breaking of the coherence, 
averaging of oscillations, experiments with the beams of mass eigenstates and measurements of the 
survival (rather than transition) probabilities are the key elements of the  
method\rlap{.}\,\cite{yasaman}  

\subsection{\bf Neutrinos from SN1987A}
\label{subsec:sol}

After confirmation of the LMA-MSW solution we can definitely say that the
effect of  flavor conversion has already been observed  in 1987. 
One must take into account the conversion effects in analysis of 
SN1987A\cite{sn87a} and future  supernova  neutrino data. 

\begin{figure}[h!]
\begin{center}
\hspace{-0.1cm} \epsfxsize8.5cm\epsffile{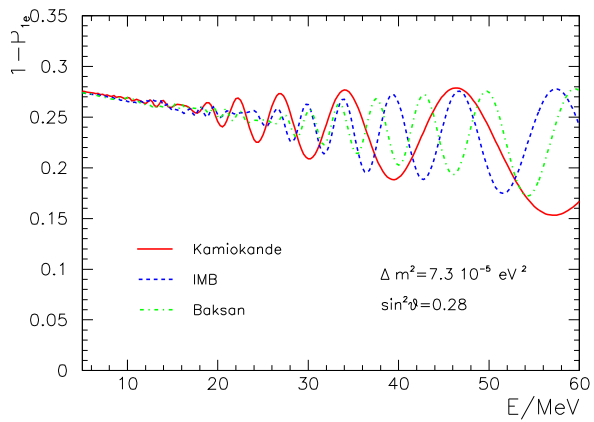}
\vskip -0.5cm
\caption{The permutation factor $\bar p = 1 - P_{1e}$ as a function of neutrino energy for 
Kamiokande II, IMB and Baksan detectors.$^{26}$ }
\label{sn87}
\end{center}
\end{figure}

In terms of the original fluxes of the electron and muon antineutrinos,  
$F^0(\bar \nu_e)$ and $F^0(\bar \nu_{\mu})$,  the electron antineutrino 
flux at the detector can be written as 
\begin{equation}
F(\bar \nu_e) = F^0(\bar \nu_e) + \bar{p} \Delta F^0,  
\label{flu}
\end{equation}
where $\Delta F^0 \equiv F(\bar \nu_{\mu}) -  F(\bar \nu_e)$, 
and $\bar{p}$ is the permutation factor. In  assumptions of the  normal mass hierarchy 
(ordering) and the absence of new neutrino states, $\bar{p}$ can be calculated 
precisely: $\bar{p} = 1 - P_{1e}$,  where $P_{1e}$ is  
the probability of $\bar{\nu}_1 \rightarrow \bar{\nu}_e$ transition 
inside the Earth\rlap{.}\,\cite{DS,LS-87} 
It can be written as $\bar{p} = \sin^2 \theta_{12} + f_{reg}$, where $f_{reg}$ describes  the effect of 
oscillations (regeneration of the $\bar{\nu}_e$ flux) inside the Earth. 
Due to the difference in distances traveled by neutrinos to 
Kamiokande, IMB and Baksan detectors inside the Earth:   
4363~km, 8535~km and 10449~km correspondingly,  the permutation factors differ   
for these detectors (Fig.~\ref{sn87}).  
The Earth matter effect can partially explain the difference between the 
Kamiokande and the IMB spectra of events\rlap{.}\,\cite{LS-87} 

In contrast to $\bar p$,  the original fluxes, and consequently $\Delta F^0$,  
are not well known, and one can not make precise predictions of the flux modification 
in Eq.~(\ref{flu}). 
\vskip -1cm
For the inverted mass hierarchy and  $\sin^2 \theta_{13} > 10^{-5}$ 
one would get a stronger permutation,    
$\bar{p} = 1$, and therefore a harder $\bar\nu_e$ spectrum,
as well as  the  absence of the Earth matter effect. This is disfavored by  the 
data\rlap{,}\,\cite{sn-inv}  
though in view of small statistics and uncertainties in the original fluxes  
it is not possible to make a firm statement. 

\subsection{\bf Absolute Scale of Mass}
\label{subsec:abs}

\begin{figure}[h!]
\begin{center}
\hspace{-0.1cm} \epsfxsize7.5cm\epsffile{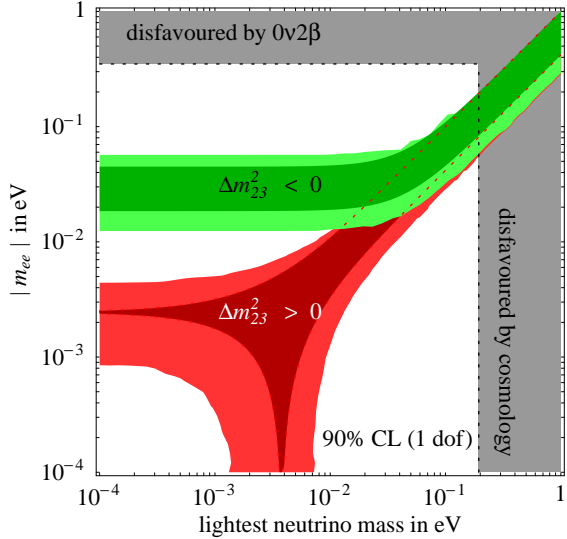}
\caption{The 90\% CL range for $m_{ee}$ as a function of the lightest neutrino mass 
for the normal ($\Delta m_{23}^2 > 0$) and inverted ($\Delta m_{23}^2 < 0$) mass hierarchies.$^{28}$ 
The darker regions show how the allowed range for the present best-fit values of the parameters 
with negligible errors.    
}
\label{bb}
\end{center}
\end{figure}

{}From the oscillation results  we can put only a lower limit on the heaviest 
neutrino mass: 
\begin{equation}
m_h  \geq \sqrt{\Delta m^2_{13}} > 0.04~ {\rm eV}, 
\label{eq:ab1}
\end{equation}
where $m_h = m_3$ for the normal mass hierarchy,  and $m_h = m_1 \approx m_2$ for the 
inverted hierarchy. 
The neutrinoless double beta decay is determined by the combination 
\begin{equation}
m_{ee} = |\sum_k U_{ek}^2 m_k e^{i\phi(k)}|, 
\label{eq:it}
\end{equation}
where $\phi(k)$ is the phase of the $k$ eigenvalue. 
Figure~\ref{bb} summarizes the present knowledge of the absolute mass scale.  
Shown are  the allowed regions in the plane of  $m_{ee}$ probed by $\beta\beta_{0\nu}$ decay 
and  the mass of 
lightest neutrino  probed by the direct kinematical methods and cosmology. 
The best present bound on $m_{ee}$ is given by the Heidelberg-Moscow experiment: 
$m_{ee} < 0.35 - 0.50$ eV\rlap{,}\,\cite{HM-neg} part of collaboration claims evidence of a positive signal\rlap{.}\,\cite{HM-pos}   
Interestingly, the present  double beta decay measurements and 
cosmology have similar sensitivities  $m_{ee} \sim m_1 \sim (0.2 - 0.5)$ eV.
The latter  corresponds to the degenerate mass spectrum:    
$m_{1} \approx m_2 \approx m_3 \equiv m_0$. 
Analyses of cosmological data (with WMAP)  result in  the 95\% C.L. 
upper bounds $m_0 < 0.23$ eV\rlap{,}\,\cite{cosm}  
$m_0 < 0.6$ eV\cite{cosm1} and $m_0 < 0.34$ eV\rlap{.}\,\cite{cosm2} 
Independent analysis which includes the X-ray galaxy cluster data  gives 
non-zero  value $m_0 = 0.20 \pm 0.10$ eV\rlap{.}\,\cite{cosm3}


Future improvements of the upper bound on $m_{ee}$ have the potential to distinguish between the 
hierarchies: According to Fig.~\ref{bb},  if the bound $m_{ee} < 0.012$ eV is established, 
the inverted hierarchy will be excluded at 90 \% C.L.. 

\subsection{\bf LSND}

The situation with this ultimate neutrino anomaly\cite{LSND} 
is really dramatic: all suggested 
physical (not related to the LSND methods) 
solutions are strongly or very strongly disfavored now. 
At the same time, being confirmed, the oscillation interpretation of  
the LSND result  may change our understanding the  neutrino 
(and in general fermion) masses. 

A recent analysis performed by the KARMEN collaboration\cite{KARMEN} 
has further disfavored a scenario\cite{BaP} 
in which the $\bar{\nu}_e$ appearance  is explained by the 
anomalous muon decay  $\mu^+ \rightarrow \bar{\nu}_e \bar{\nu}_{i} e^+$  
$(i = e , \mu, \tau)$. 

The CPT-violation scheme\cite{BLyk}  with different mass spectra 
of neutrinos and antineutrinos 
is disfavored by the atmospheric neutrino data\rlap{.}\,\cite{stru} No compatibility of LSND and 
``all but LSND" data have been found below $3\sigma$\rlap{.}\,\cite{coCPT} 

The main problem of the (3 + 1) scheme with 
 $\Delta m^2 \sim 1$ eV$^2$  is that the predicted LSND signal,
which is consistent with the results  of other short base-line experiments
(BUGEY, CHOOZ, CDHS, CCFR, KARMEN) as well as the atmospheric neutrino data,  is too small:
the $\bar{\nu}_{\mu} \rightarrow \bar{\nu}_e$ probability is about $3\sigma$ below the 
LSND measurement.

Introduction of the second sterile  neutrino 
with $\Delta m^2 >  8$ eV$^2$ may help\rlap{.}\,\cite{PS31}
It was shown\cite{sorel}  that a new neutrino with
$\Delta m^2 \sim 22$ eV$^2$ and mixings $U_{e5} = 0.06$, $U_{\mu5} = 0.24$
can enhance the predicted LSND signal by (60\,--\,70)\% .
The (3 + 2) scheme has, however, problems with cosmology and astrophysics. 

\begin{figure}[h!]
\begin{center}
\hspace{-0.1cm} \epsfxsize8cm\epsffile{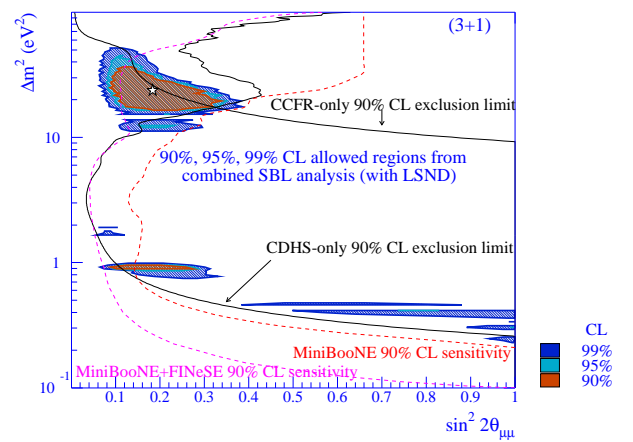}
\caption{The allowed regions of parameters of the (3 + 1) scheme, 
$\Delta m^2_{14}$ and $\sin^2 2\theta_{\mu \mu} \approx 4|U_{\mu 4}|^2$,  
at different confidence levels. Shown are the 90\% sensitivity  limits of the 
MiniBooNE and  the MiniBooNE+FINeSE experiments. 
}
\label{boon}
\end{center}
\end{figure}

The generic prediction of (3 + n)  schemes is the $\nu_{\mu}$ oscillation 
disappearance at the  level of existing  
upper bouds from CDHS\rlap{,},\cite{CDHS} CCFR\rlap{,},\cite{CCFR} and NOMAD\cite{NOMAD} experiments. 
New searches of $\nu_{\mu}$ disappearance are being performed by the MiniBooNE 
experiment\cite{mini} and planned by the proposed experiment FINeSE\cite{fine} 
(see Fig.~\ref{boon}, where the sensitivity region of these searches  
is shown\cite{fine}).  

The combination of the two described solutions, namely the $3 + 1$ 
scheme with CPT-violation,  has been considered\rlap{.}\,\cite{barger3}

\subsection{\bf Known and Unknown}

Information described in the previous sections  
can be summarized in the following way. 

1. The observed ratio  of the mass squared differences, 
$\Delta m^2_{12}/\Delta m^2_{23} = 0.01 - 0.15$,  
implies that there is no strong hierarchy of masses: 
\begin{equation}
\frac{m_2}{m_3} >  \sqrt{\frac{\Delta m^2_{12}}{\Delta m^2_{23}}} = 0.18^{+ 0.22}_{-0.08}. 
\label{eq:hie1}
\end{equation}
For charge leptons the corresponding ratio is 0.06.

2. There is the bi-large or large-maximal mixing between the neighboring families
(1 - 2) and (2 - 3). Still  rather significant deviation of the 2-3 mixing 
from the maximal one is possible. 

3. Mixing between remote (1-3) families is weak.  

Several key elements are unknown yet leading to a variety of possible interpretations. 

Knowledge of the absolute mass scale, type of mass spectrum, and  type of mass 
hierarchy is of the highest priority. 
The 1-3 mixing has important phenomenological consequences;  
its value is a test of the mechanisms of the 
lepton mixing enhancement. 
The CP-violating  Majorana phases are extremely important for the structure 
of neutrino mass matrices. 
Deviations of the 2-3 and 1-2 mixings from maximal values play a crucial role 
in understanding the origins of neutrino masses. 
The existence of new neutrino states (their search  should be a permanent 
item in the scientific agenda) may change completely our approaches to the 
underlying theory. 
 
These are phenomenological and experimental questions we will deal with during the next 
20\,--\,30 years.

\section{Open Theoretical Questions}

What does all this (results on neutrino masses and mixing) mean? 

Among old, still open, questions are the following:   
Why are neutrino masses so small  
in comparison with the charged lepton and quark masses? 
What is the origin of neutrino mass? Is it the same as the one for quarks and 
charged leptons? 

What are the relations between  neutrino masses and other mass/energy scales in nature,  
e.g. the scale of cosmological constant or dark energy?

Why is the lepton mixing large? Why is it so different from quark mixing? 
Is the 2-3 mixing exactly maximal? 
What are the relations between different mixing angles (if any)?
How is the observed pattern of lepton mixing is generated? 

In the quark sector the smallness of  mixing is related to the
strong mass hierarchy. 
What are the relations between the lepton masses and lepton mixing? 

Do neutrinos show certain flavor or horizontal symmetry? 
If so, is this symmetry consistent with the 
pattern of quark masses and mixing? 

Are the results of neutrino masses and lepton mixing consistent with 
the quark-lepton symmetry and  Grand Unification? 

If new light sterile neutrinos exist 
what is their nature and why are they light? 

What are the implications of the neutrino results for GUT, SUSY, models with extra dimensions, and strings? 
{\it Vice versa:} what can these beyond the SM theories tell us about neutrinos? 

\section{Bottom-Up}

One can try the ``top-down" approach confronting immediately a  proposed  
model with experimental results.   
Inversely, to get some hints in answering the above questions,  
it may be worthwhile to try to move bottom-up.

\subsection{\bf Neutrino mass matrix}

There are several steps  in the bottom-up approach.   

1). Take the  results on $\Delta m^2_{ij}$, $\theta_{ij}$, $m_{ee}$,  {\it etc.}.   

2). Reconstruct the neutrino mass matrix in the flavor basis   
(where the charge lepton mass matrix is diagonal)  assuming also that neutrinos 
are Majorana particles.  Notice that the mass matrix unifies information contained in masses 
and mixing angles and this may provide some more  hints toward the underlying theory. 

3). Identify the symmetry basis (which may differ from the flavor basis)    
and the symmetry scale. Take into account the renormalization 
group effects. 

4). Identify the symmetry (as well as mechanism of symmetry violation, if needed) 
and underlying dynamics. 

Let us make  the first step in the bottom-up approach.   
The mass matrix in the flavor basis can be written as 
\begin{equation}
m = U^* m^{diag} U^+,
\label{eq:mass}
\end{equation}
where $U = U(\theta_{ij}, \delta)$ 
is the mixing matrix, $\delta$ is the Dirac CP-violating phase,   
and 
\begin{equation}
m^{diag} = diag (m_1 e^{-2i\rho},~ m_2,~ m_3 e^{-2i\sigma}).  
\label{eq:mass}
\end{equation}
Here $\rho$ and $\sigma$ are the Majorana phases. 
The mass eigenvalues equal $m_2 = \sqrt{m_1^2  + \Delta m^2_{12}}$, and
$m_3 = \sqrt{m_1^2  + \Delta m^2_{13}}$. 

\begin{figure}[h!]
\begin{center}
\hspace{-0.1cm} \epsfxsize8cm\epsffile{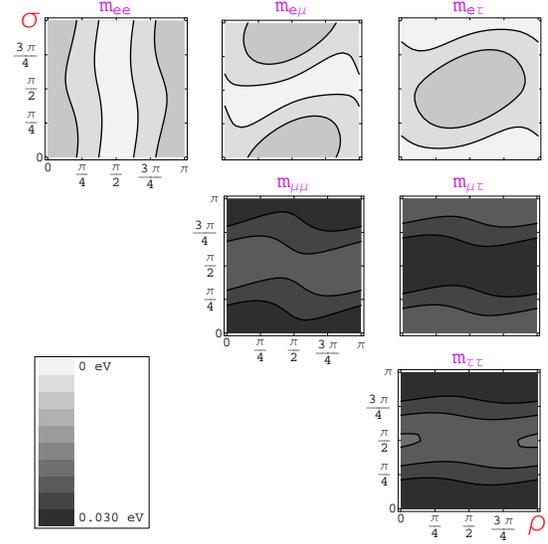}
\caption{The Majorana mass matrix for the
normal mass hierarchy: $m_3/m_2=5$, 
$m_1\approx 0.006$ eV. 
We show contours of  constant mass in 
the $\rho-\sigma$ plots for the moduli of  mass matrix elements.  
We take  for other parameters $\Delta m^2_{12}=7 \times 10^{-5}
{\rm eV}^2$, $\Delta m^2_{13}= 2.5 \times 10^{-3} {\rm eV}^2$, 
$\tan^2\theta_{12}=0.42$, $\tan\theta_{23}=1$, $\sin \theta_{13}=0.1$, and $\delta=0$.}
\label{norm5}
\end{center}
\end{figure}

\begin{figure}[h!]
\begin{center}
\hspace{-0.1cm} \epsfxsize8cm\epsffile{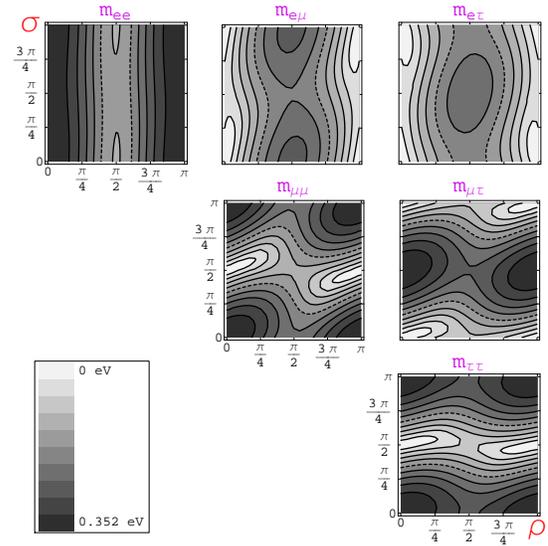}
\caption{The same as in Fig. \ref{norm5} for the quasi-degenerate spectrum:
$m_3/m_2 =  1.01$, $m_1\approx 0.35$ eV.}
\label{deg1}
\end{center}
\end{figure}

\begin{figure}[h!]
\begin{center}
\hspace{-0.1cm} \epsfxsize8cm\epsffile{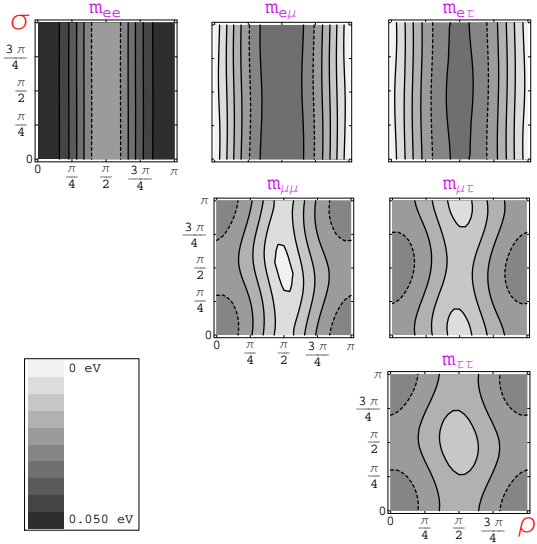}
\caption{The same as in Fig.~\ref{norm5} for the inverted mass hierarchy: 
$m_3/m_2 = 0.1$, $m_3\approx 0.005$ eV.
}
\label{inv01}
\end{center}
\end{figure}

The results of reconstruction of the mass matrix\cite{alta,matrix} are shown in 
Figs.~\ref{norm5},~\ref{deg1}, and~\ref{inv01} 
as the   $\rho - \sigma$ plots for the absolute values of the   
6 independent matrix elements.  They correspond to  three 
extreme cases: normal mass hierarchy, quasi-degenerate spectrum  
and inverted mass hierarchy. 
The figures illustrate a variety of possible structures. In particular, 
for the normal mass hierarchy (Fig.~\ref{norm5}) there is clear structure with the 
dominant $\mu - \tau$ block.  
Interesting parameterizations of the mass matrix 
(up to an overall mass factor) are 
\begin{equation}
\left(
\begin{tabular}{lll}
0 & 0 & $\lambda$ \\
0 & 1 & 1\\
$\lambda$ & 1 & 1
\end{tabular}
\right),~~~~~ 
\left(
\begin{tabular}{lll}
$\lambda^2$ & $\lambda$ & $\lambda$\\
$\lambda$ & 1 & 1\\
$\lambda$ & 1 & 1\\
\end{tabular}
\right), 
\label{eq:nhier}
\end{equation}
where $\lambda \sim 0.2$. Also the matrix similar to the first one 
in Eq.~(\ref{eq:nhier}) with $m_{12} \sim \lambda$ and  $m_{13} \approx 0$ is possible. 

In the case of a quasi-degenerate spectrum,  the interesting dominant structures are 
\begin{equation}
\left(
\begin{tabular}{lll}
1 & 0 & 0\\
0 & 1 & 0\\
0 & 0 & 1\\
\end{tabular}
\right),~~~~~~
\left(
\begin{tabular}{lll}
1 & 0 & 0\\
0 & 0 & 1\\
0 & 1 & 0\\
\end{tabular}
\right). 
\label{eq:degen}
\end{equation}
These matrices are realized for values of  phases in the corners of the plots: 
$\rho, \sigma = 0, \pi$ (the first matrix) or at $\rho = 0, \pi$, 
$\sigma = \pi/2$ (the second one) which  corresponds to definite CP-parities of 
the mass eigenstates. 
Also the ``democratic" structure with equal moduli of elements is possible for the non-trivial 
values of phases\rlap{.}\,\cite{demo} 
Changing the phases one can get any 
intermediate structure between those in Eqs.~(\ref{eq:nhier}) and (\ref{eq:degen}). 

In the case of the inverted hierarchy, generically the $ee$- element is not small. 
Among  interesting examples are
\begin{equation}
\left(
\begin{tabular}{lll}
0.7 & 1 & 1\\
1 & 0.1 & 0.1\\
1 & 0.1 & 0.1\\
\end{tabular}
\right), ~~~~~~~
\left(
\begin{tabular}{lll}
1 & $0.1$ & $0.1$\\
$0.1$ & 0.5 & 0.5\\
$0.1$ & 0.5 & 0.5 \\
\end{tabular}
\right). 
\label{eq:ihier}
\end{equation}

In the SM and MSSM the renormalization group effects do not change the 
structure of the mass matrix: the corrections to a given element are 
proportional to the element itself. Furthermore, the corrections are  
small even in the SUSY case (below $0.1\%$). 
So, unless some new interactions exist,  the mass matrix determined at low energies  
does not change structure when running up to 
the scale where the corresponding mass operators are formed or up to the  symmetry scale.  

In contrast to the matrix structure, 
the radiative corrections are important for the oscillation observables in the case of partially 
or quasi degenerate mass spectra.   

Scanning the $\rho-\sigma$ plots shown in 
Figs.~\ref{norm5},~\ref{deg1}, and~\ref{inv01},  one can make the 
following observations. 

1). A large variety of different structures is still possible, depending strongly on 
the unknown $m_1$, type of mass hierarchy and  Majorana phases. The dependence on $\sin \theta_{13}$ and $\delta$ 
is  weak. 

2). Generically the hierarchy of elements is not strong: within 1 order of magnitude. 
At the same time matrices with one or two exact zeros are not excluded\rlap{.}\,\cite{zero} 

3). Matrices  are possible with: 

- dominant (i) diagonal elements $(\sim I)$, (ii)  $\mu \tau$-block, 
(iii) $e$-row elements, (iv) $ee-, \mu\tau-, \tau\mu-$ elements (triangle structure), 

- democratic structure,  

- flavor alignment, 

- non-hierarchical structures with all elements of the same order, 

- flavor disordering, 

- zeros at different places, and

- equalities of various element. 

4). Typically,  the hierarchical structures appear for the Majorana phases near 0 , $\pi/2$, or 
$\pi$.

5). Matrices can be parameterized in terms of powers of small parameter $\lambda = 0.2 - 0.3$
consistent with the Cabibbo mixing. 

In a significant part of the parameter space the matrix does not show any regularities, and 
relative values of its elements appear as random numbers which spread within one order of 
magnitude. This supports 
the idea of ``Anarchy"\rlap{.}\,\cite{anarc,anarc1} Consideration of the  anarchy of elements is  
a test of the possible complexity of the neutrino mass matrix. 
The case of anarchy can be imitated if 
neutrino masses have several different contributions,  and even if  each of them  
has well defined structure or symmetry, the sum may show up as a matrix with disorder. 
In this connection one can consider representations of the mass matrix as the sum of 
matrices, given in Eq.~(\ref{eq:degen}), as well as the democratic matrix with certain coefficients\rlap{.}\,\cite{decomp} 

What is more fundamental: oscillation observables or neutrino mass matrix in some basis? 
The answer may depend on  the type of mass spectrum. In the case of hierarchical spectrum 
the observables are visibly  imprinted into the structure of the mass matrix. In contrast, 
for the quasi-degenerate spectrum they are just very small perturbations of the dominant structure 
which is determined by the non-oscillatory parameters: the absolute mass scale and 
the  Majorana CP-violating phases. Then the oscillation parameters 
can  be a result of interplay of some small,  
in particular, radiative corrections. 

In the case of Majorana neutrinos, the elements 
of the mass matrix are physical parameters: they can be immediately measured in the 
neutrinoless beta decay and, in principle,  in other similar processes.  
In practice, it is not possible to reconstruct   the mass matrix  from experiment completely. 
Even in the most optimistic case the phase $\sigma$ will be undetermined and 
Figs.~\ref{norm5},~\ref{deg1}, and~\ref{inv01}, give an idea of the remaining uncertainty. Only in 
the case of the g mass hierarchy does the dependence on $\sigma$ disappear. 
The hope is that even without complete reconstruction of the mass matrix  
we will be able to uncover the underlying physics. 

\subsection{\bf Neutrino Mass and Horizontal Symmetry}

Do the results on neutrino masses and mixing indicate certain regularities or symmetry? 
Can the dominant structures of the mass matrix  be explained by a symmetry with 
the sub-dominant elements appearing as a result of violations of the symmetry? 
Is the neutrino mass matrix consistent with symmetries suggested for quarks? 

The following symmetries have been considered.

1). $L_e - L_{\mu} - L_{\tau}$\rlap{.}\,\cite{emt} This symmetry supports,  in particular, 
the structure with an inverted mass hierarchy. However, the rather large element 
$m_{ee}$  (Fig. \ref{inv01}) shows strong violation of this symmetry. 

2). Discrete symmetries: $A_4$\rlap{,}\,\cite{A4} $S_3$\rlap{,}\,\cite{S3} 
$Z_4$\rlap{,}\,\cite{Z4}, and $D_4$\rlap{.}\,\cite{D4}  
They  reproduce successfully the dominant structures  
in Eq.~(\ref{eq:degen}) as well as the ``democratic" matrix. 

However, both classes of symmetries 1) and 2) typically treat quarks and leptons differently.

3). $U(1)$\rlap{.}\,\cite{u1} In the  Froggatt-Nielsen context\cite{FN} this symmetry can describe mass 
matrices of both 
quarks and leptons. However, the claimed  predictability of this approach can 
be questioned: the $U(1)$ charges should be considered as 
discrete free parameters. Furthermore, precise description of data usually requires   
coefficients  (prefactors) of the order 1  (1/2 - 2) in front of powers of the expansion parameter. 
The outcome is that the   mixing pattern  depends substantially on values of these unknown prefactors.

4). $SU(2)$\rlap{,}\,\cite{su2}  $SO(3)$\rlap{,}\,\cite{so3}, and $SU(3)$\cite{su3} require a complicated 
Higgs sector to break the symmetry.  
Often models are too restrictive and predictions are on the borders of allowed regions.

The question is still open. 
Different symmetries are consistent with 
the neutrino data.  
But realizations of these symmetries in specific models  are not simple. The hope is 
that future neutrino data (better knowledge of the mass matrix) can discriminate 
among possibilities.  

\section{How We May Go...}

\subsection{\bf Neutrality and Mass}

In answering the questions of Sec. 3 one can  
implement the ``minimalistic" approach,  
that is, to try to relate features of the neutrino masses 
and mixings with already known differences of characteristics of  
neutrinos  and other fermions. 

The main feature of neutrinos is  neutrality: 
\begin{equation}
Q_{\gamma} = Q_c  = 0. 
\label{eq:mass}
\end{equation}
It leads to the following possibilities: 

\begin{itemize} 

\item
neutrinos can be Majorana particles;  

\item
they  can mix with singlets of the SM symmetry group; and

\item
the right-handed components (RH), if they exist, are singlets of 
$SU(3)\times SU(2)\times U(1)$. So, their masses are unprotected by the symmetry  
and therefore can be large.  

\end{itemize}

In turn,  properties  of the RH components open two other possibilities. The RH neutrinos can:  

\begin{itemize}

\item
have large Majorana masses: $M_R \gg V_{EW}$ 
(which leads to the see-saw); and

\item
propagate in (large, or warped,  or infinite) extra dimensions, 
or be located on the ``hidden" (not ours) brane in contrast to other fermions.  

\end{itemize}

Introduction of the RH neutrino has a number of attractive features\rlap{,}\,\cite{right} in particular, 
it  allows one to extend the electroweak symmetry to the gauged 
$SU(2)_L \times SU(2)_R \times U(1)_{B-L}$. 

Is this enough to explain the properties of the mass spectrum and mixings? 

\subsection{\bf Effective Operator}

Suppose the SM particles are the only light degrees of freedom.
Then at low energies (after integrating out the heavy degrees of freedom) 
one can get the operator:\,\cite{eff}
\begin{equation}
\frac{\lambda_{ij}}{M} (L_i H)^T(L_j H), ~~~ i,j = e, \mu, \tau ,
\end{equation}
where $L_i$ is the lepton doublet,  
$\lambda_{ij}$  are the dimensionless couplings and ${M}$ is the
cut-off scale. After EW symmetry breaking it generates the neutrino
masses
\begin{equation}
m_{ij} = \frac{\lambda_{ij} \langle H \rangle^2}{M}. 
\end{equation}

For $\lambda_{ij} \sim 1$ and $ M = M_{Pl}$ we find $m_{ij} \sim 10^{-5}$~eV\rlap{.}\,\cite{planck}
Three important conclusions immediately follow from this 
consideration.

1). The Planck scale (gravitational) interactions  are not enough to
generate the observed values of the masses. So, new scales of physics
below $M_{Pl}$ should exist.

2). Contributions to the neutrino masses of the order $\sim 10^{-5}$ eV are
still relevant for phenomenology. Sub-dominant structures of the
mass matrix can be generated by the Planck scale interactions\rlap{.}\,\cite{BerV}

3). The neutrino mass matrix can get observable contributions
from all possible energy/mass scales from the EW scale to the Planck
scale. As a consequence, the structure of the mass matrix can be rather complicated.

\subsection{\bf See-saw}

The see-saw (type I) mechanism\cite{sees} implements the neutrality 
in full strength (Majorana nature, heavy RH components). 
Introducing the Dirac mass matrix, $m_D = Y v_{EW}$, 
where $Y$ is the matrix of Yukawa couplings and 
$v_{EM}$ is the electroweak VEV, we have 
\begin{equation}
m = - m_D^T M_R^{-1} m_D~~~~~   (type I). 
\label{eq:seesaw}
\end{equation}

If the $SU(2)$ triplet, $\Delta_L$, exists which develops a VEV $\langle \Delta_L \rangle$, 
the left-handed neutrinos  can get a direct mass $m_L$ via the interaction $f_{\Delta} L^T L \Delta_L$. 
If $\Delta_L$ is very heavy, it can develop   the induced VEV from interactions with a doublet: 
$\langle \Delta_L \rangle = v_{EW}^2/M$.  So that 
\begin{equation}
m_L = f_{\Delta}\frac{v_{EW}^2}{M}~~~~~(type II),     
\label{eq:mass}
\end{equation}
and   here we deal with the see-saw of VEV's\rlap{.}\,\cite{sees2} 

In $SO(10)$ with $126_H$-plet of Higgses we have  $M_R = f v_R$, where $f$ is the Yukawa coupling 
of the matter 16-plet with 
$126_H$ and $v_R$ is the VEV of the $SU(5)$ singlet component of $126_H$. Now   
$f_{\Delta} = f$, and  the general mass term which contains  both types of contributions 
can be written as 
\begin{equation}
m = \frac{v_{EM}^2}{v_R} (f \lambda - Y^T f^{-1} Y).  
\label{eq:mass}
\end{equation}
Here $\lambda$ is the coupling of 10- and 126-plets. 
According to this expression  the flavor structure of the two contributions may partially 
correlate.

The number of RH neutrinos can differ from 3. Two possibilities have been explored: 

``{\it ... less than 3~}": which corresponds  to the $3\times2$ see-saw 
in the case of two RH neutrinos\rlap{.}\,\cite{less} 
Such a possibility can be realized in the limit when one of the RH neutrinos is very heavy: 
$M \sim M_{Pl}$, being, e.g. unprotected by the $SU(2)_H$ horizontal symmetry. 
It leads to one exactly massless LH neutrino and smaller number of free parameters. 

One can further reduce the  number of unknown 
parameters postulating zeros in  the Dirac matrix 
$m_D$\rlap{.}\,\cite{zero} This can lead to the predictions for $\sin \theta_{13}$, 
$m_{ee}$, as well as for relations between  $\delta$ and the phase responsible for leptogenesis.

``{\it ... more than 3~}": additional singlets of the SM may not be related 
to the family structure.  Alternatively, three additional singlets, $S$, 
which belong to families,  
can couple to  the RH neutrinos. In the latter  case the double see-saw can be realized\rlap{.}\,\cite{dsees}
In the basis $(\nu, \nu^c, S)$,   the mass matrix may have the form 
\begin{equation}
\left(
\begin{tabular}{lll}
0 & $m_D$ & 0\\
$m_D^T$ & 0 & M\\
0 & $M^T$ & $\mu$ \\
\end{tabular}
\right) 
\label{eq:ihier}
\end{equation}
which leads to the light neutrino masses: 
\begin{equation}
m = - m_D^T (M^{-1})^T \mu M^{-1} m_D . 
\label{eq:mass}
\end{equation}
Two interesting limits are: 
(i)  $\mu \ll M$, it  allows one to reduce all high mass  scales  
for the same values of the light neutrino masses,  (ii) 
$\mu \gg M$,  {\it e.g.}
$\mu =  M_{Pl}$, and $M = M_{GU}$: in this case  the intermediate mass scale,  
$M_{GU}^2/M_{Pl} =  10^{12} - 10^{14}$ GeV  for the masses of RH neutrinos can be obtained.

\subsection{\bf Grand Unification and Neutrino Mixing}

GU theories provide a large mass scale comparable to the scale of  RH neutrino masses\rlap{.}\,\cite{GUT} 
Furthermore, one can argue that GUT + see-saw can naturally lead to the large lepton mixing 
in contrast to the quark mixing.  
The arguments go like this: 

1. Suppose that all quarks and leptons of a given family  are in a single multiplet $F_i$ 
(as 16 of SO(10)). 

2. Suppose that all Yukawa couplings are of the same order thus 
producing matrices with generically large mixing.

3. If the Dirac masses are generated by an unique Higgs multiplet, say $10_H$ of SO(10),  
the mass matrices of the up and down components of the weak doublets have 
identical structures, and so,  will be diagonalized by  the same rotations. As a result:

- no mixing appears for quarks, and

- masses of up and down components will be equal to each other
(this  needs to be corrected).

4. In contrast to other fermions,  the RH neutrinos acquire Majorana masses via the  additional Yukawa couplings  
(with $126_H$ of SO(10)).  

5. If those (Majorana type) Yukawa couplings are also of the generic form, 
they produce $M_R$ with large mixing which 
leads then to large lepton mixing. 

The problem of this scenario is the strong hierarchy of the quark and lepton masses. 
Indeed, taking the neutrino Dirac masses as $m_D = diag(m_u, m_c, m_t)$ in a spirit of GU, we find 
that for generic $M_R$ the see-saw type I  
formula (\ref{eq:seesaw}) produces strongly hierarchical mass matrix with small  mixings. 
Possible solutions are: 

\begin{enumerate}

\item
a special structure of $M_R$ which compensates the  strong hierarchy in $m_D$;

\item
a substantial difference in the  Dirac matrices of quarks and leptons: 
$m_D(q) \neq m_D(l)$; or

\item
a type II see-saw for which there is no relation to  $m_D$. 

\end{enumerate}

In what follows we will comment on these three possibilities. 

{\it See-Saw enhancement of mixing\rlap{.}\,\cite{senhan}} Can the same mechanism (see-saw) which explains the smallness of 
the neutrino mass also explain the large lepton mixing? So,  that the large  mixing appears as an artifact of 
the see-saw?

The idea is that  due to the (approximate) quark-lepton symmetry, the Dirac mass matrices 
of the quarks and leptons have 
the same (similar) structure   
$m_D \sim m_{up}$,  $m_l \sim m_{down}$ leading to small mixing in the Dirac sector. However, 
the special structure of $M_R$ (which has no analogue in the quark sector) leads to an enhancement of 
lepton mixing. Two different possibilities have been found:\,\cite{} 

\begin{itemize}

\item
strong (nearly quadratic) hierarchy of the RH neutrino masses: 
$M_{iR} \sim (m_{i up})^2$; and

\item
strong interfamily connection (pseudo Dirac structures) like 
\begin{equation}
M_R = \left(
\begin{tabular}{lll}
A & 0 & 0\\
0 & 0 & B\\
0 & B & 0 \\
\end{tabular}
\right).
\label{eq:}
\end{equation}

\end{itemize}

In the three neutrino context both possibilities can be realized simultaneously, 
so that the pseudo Dirac structure leads to maximal 2-3 mixing, whereas the strong hierarchy 
$A \ll B$ enhances the 1-2 mixing. 

\begin{figure}[h!]
\begin{center}
\hspace{-0.65cm} \epsfxsize8.5cm\epsffile{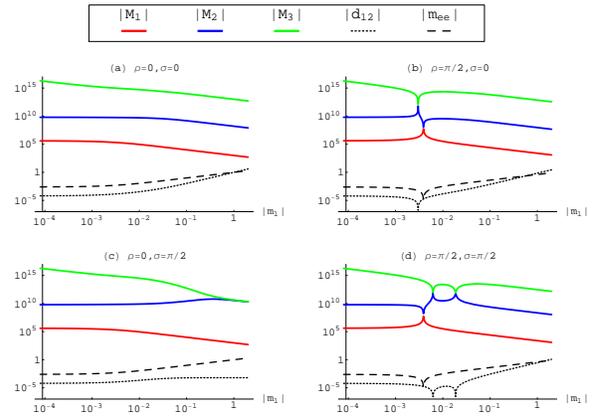}
\caption{The masses of the RH neutrinos in GeV as functions of the lightest neutrino 
mass $|m_1|$ in eV (solid lines) for different values of the Majorana phases of light neutrinos.$^{75}$  
We take $\sin\theta_{13} = 0$ and the best-fit values of other oscillation parameters. 
Shown is also the dependence of $m_{ee}$ (in eV) on $|m_1|$ (thin dashed line). 
}
\label{rhmass}
\end{center}
\end{figure}

In Fig.~\ref{rhmass}  we show dependences of  the RH neutrino masses 
reconstructed from the low energy data on  the lightest neutrino mass 
for different values of the Majorana phases. 
According to this figure 

1). In the largest part of the parameter space ($m_1$, $\rho$, $\sigma$)  there is a very 
strong (4 - 5 orders of magnitude) mass hierarchy of the RH neutrinos.  

2). The lightest mass  is typically below $10^{5}$ GeV, thus 
strongly violating the lower bound on the mass 
from the condition of successful leptogenesis: $M_1 > 4 \times 10^8$ GeV\rlap{.}\,\cite{buch} 

3). At certain points the level crossings occur. At these points 
(i) there is a strong degeneracy of  mass eigenstates: in particular, $M_1 = M_2$, 
(ii) the lightest mass can be as large as $10^8$ GeV and 
(iii) the lepton asymmetry can be resonantly enhanced\cite{reslep} up to the required value.

{\it Large mixing and type II see-saw}. In general, the structure of neutrino mass 
matrix generated  by the type II (triplet) see-saw is not related to structures 
of matrices of other fermions. 

In some particular cases, however, the relations can appear leading to 
interesting consequences.  
In the SO(10) model the $126_H$ Higgs multiplet can play a double role: 
(i) generate neutrino masses  $m_L = Y_{126} v_{\Delta}$, where $v_{\Delta}$ is 
the VEV of the SU(2) triplet in $126_H$; and (ii) give  contributions  
to the quark and charged lepton masses  (if doublets contained in 126 get VEV's) 
reproducing the Georgi-Jarlskog mass relation for the first and second generations. 

Since $m_b - m_{\tau} \propto (Y_{126})_{33}$, 
the  contribution of $126_H$ destroys the  $b - \tau$ unification 
unless  $(Y_{126})_{33} \leq (Y_{126})_{23}$. The latter  leads to large 
(but not necessarily maximal) 2-3 lepton mixing. 
In this context, the $b - \tau$ unification implies large 2-3 mixing\rlap{.}\,\cite{btau} 

In models of this type a successful leptogenesis is possible with participation of the 
scalar triplet\rlap{.}\,\cite{HS}  
The model has been generalized also to 3 generations, leading typically  to 1-2 mixing 
at the larger side of the allowed region\rlap{.}\,\cite{3gen} 

{\it Single RH neutrino dominance\rlap{.}\,\cite{sdom}} The 
large neutrino mixing and relatively strong mass hierarchy 
implied by the solar and atmospheric neutrino data can be reconciled if only  
one RH neutrino gives the dominant contribution to the see-saw. 
(This leads to the submatrix of $m_L$ with nearly zero determinant.) 
There are two different realizations of this possibility. 
In one case the large mixing originates  from the large mixing in  the 
Dirac neutrino mass matrix $m_D$: 
two LH neutrinos have nearly equal couplings to the  dominating RH component. 
Suppose that  $(m_D)_{23} \approx (m_D)_{33} = m$, $(m_D)_{13} = 
\lambda m$ ($\lambda \approx 0.2$) and all other elements of $m_D$ are much smaller. Then if  
only $(M^{-1})_{33}$ is large in the inverted matrix, the  
see-saw leads to the mass matrix which reproduces  the second structure in Eq.~(\ref{eq:nhier}). 

In another version, the  dominance is realized when two RH neutrinos are much heavier  
than the third (dominating) one and no large mixing in $m_D$ appears. 
This is equivalent to the strong mass hierarchy case 
of the see-saw enhancement mechanism. 
A realization requires  $(m_D)_{22} \approx (m_D)_{23} \ll (m_D)_{33}$,  
and  $(M^{-1})_{22}$ being the dominant element.

{\it Lopsided models\rlap{.}\,\cite{lops}}
Large lepton mixing follows from the charge lepton mass matrix  
which  should be non-symmetric (no left-right symmetry). 
This does  not contradict the Grand Unification: in SU(5) the LH components of leptons are unified 
with the RH components of quarks: $5 = (d^c, d^c, d^c, l, \nu)$. Therefore large 
mixing of the LH leptonic components is accompanied by  large mixing 
of the RH $d$-quarks which is unobservable. Introducing the 
Dirac mass matrix of the charged leptons with the only large elements  
$(m_l)_{33} \sim (m_l)_{23}$, one can obtain the large 2-3 lepton mixing. 
This scenario can also  be  realized in $SO(10)$, if the symmetry is broken via $SU(5)$. 
A double lopsided matrix for both large mixing is also possible. 


{\it Radiative enhancement of mixing}\rlap{.}\,\cite{rad}  The idea is 
that the difference between the  quark and lepton mixings is a result of different renormalization group 
effects. The lepton mixing is small 
(similar to quark mixing) at the GU scale but running to 
low energies leads to its enhancement. 

The main requirement of such an enhancement is that the neutrino mass 
spectrum is quasi-degenerate (and this is the key point which distinguishes quarks and leptons). 
The enhancement 
occurs when neutrinos become even more 
degenerate at low energies. 
For instance, running of the 2-3 mixing is described by 
\begin{equation}
\frac{d \sin\theta_{23}}{dt} \sim (\sin\theta_{12} U_{\tau1} D_{31} - 
\cos\theta_{12} U_{\tau2} D_{32}), 
\label{eq:evol}
\end{equation}
where $t \equiv 1/8\pi^2 log (q/M)$, $D_{ij} \equiv (m_i + m_j)/( m_i - m_j)$, and 
$m_i$ are the mass eigenvalues. The minus sign in the denominators of $D_{ij}$ plays the key 
role. 

The mechanism requires fine tuning of the initial mass splittings and radiative 
corrections. 
In principle,  in the MSSM both the 1-2 and 2-3 mixings can be enhanced in this way. 
In the SM only 1-2 mixing can be enhanced.  

The fine-tuning problem can be avoided 
if the masses are generated from the K\"ahler potential. In this case  large mixing appears as 
the infrared fixed point\rlap{.}\,\cite{casas}  

Another possible application of the radiative effects is the generation of small oscillation 
parameters like $\Delta m_{12}^2$, and $\sin\theta_{13}$\rlap{.}\,\cite{small} Again this can be realized 
only in the case of a mass spectrum with degeneracy.

\subsection{\bf How To Test the See-Saw Mechanism?}

This is the key question which  implies essentially  the test of existence of the heavy Majorana RH 
neutrinos. There are two (known) possibilities. 

1).  Leptogenesis\rlap{.}\,\cite{FY} For the hierarchical RH neutrino spectrum 
and in assumption of the type I see-saw, it gives bounds on 
(i) the mass of the lightest RH neutrino $M_{R1}$, (ii) the  effective parameter
$\tilde m_1$ which determines the washout effect.  Notice that 
the leptogenesis  probes the combination of the Yukawa couplings $(Y Y^{\dagger})_{ii}$.  

2). The RH neutrinos can produce renormalization effects above
the scale of their masses: between $M_R$ and, say,  the GUT scale.
In particular, they can renormalize the $m_b - m_{\tau}$ mass
relation\cite{radbtau} which leads to the observable effect
in the assumption of $m_b - m_{\tau}$ unification at the GUT scale.

Another possibility is that the renormalization due to RH neutrinos modifies 
masses and mixing of the light neutrinos, e.g. enhances  the lepton
mixing\rlap{.}\,\cite{massre}

\subsection{\bf SUSY See-Saw}

Additional possibilities to test the see-saw mechanism appear if SUSY is realized.
The part of superpotential relevant for the see-saw can be written as
\begin{equation}
W_{lep} = l^{cT} Y_l L H_1  + \nu^{cT} Y L H_2 +
\frac{1}{2} \nu^{cT} M_R \nu^{c}.
\label{wpot}
\end{equation}
Structures relevant for the see-saw are imprinted into 
properties  of the slepton sector. So, studying the properties of sleptons
(masses, decay rates, etc.) one can get information about  the neutrino mass generation.

Certain predictions can be made in the assumptions of
universal soft SUSY breaking masses
($m_0$, $A_0$) at high (GUT ?) scale $M_X$, and the absence of new
particles/interactions up to $M_X$.

Due to renormalization group effects  
the Yukawa couplings  (\ref{wpot}) give 
contributions to the masses of left sleptons at low 
energies:\,\cite{borz}
\begin{equation}
(m_S^2)_{ab} =   
m_a^2 \delta_{ab} - \frac{3 m_0^2 + A_0^2}{8\pi^2} 
(Y^{\dagger})_{ai}(Y)_{ib}log\left(\frac{M_X}{M_{iR}}\right).
\label{slepton}
\end{equation}
The contribution  splits masses of sleptons of different flavors and
sleptons-antisleptons as well as leading to 
mixing of sleptons (the off-diagonal terms in Eq.~(\ref{slepton})) 
which is related to the mixing of neutrinos.

In turn, these contributions to the slepton masses 
produce a number of observable effects:

1. rare leptonic decays: the one-loop mixing of sleptons of different flavors induces 
the flavor violating decays:
$\mu \rightarrow e \gamma$, $\tau \rightarrow \mu \gamma$,
$\tau \rightarrow e \gamma$;\,\cite{borz}

2. sneutrino flavor oscillations;\,\cite{flos}

3. slepton decays;\,\cite{sdec}

4. sneutrino-antisneutrino oscillations;\,\cite{sas} and

5. contribution to the electric dipole moments of charged leptons\rlap{.}\,\cite{de}

Up to a log factor these effects are determined by the combination $\sim (Y^{\dagger}Y)$.
Notice that another combination: $(Y^T  M^{-1} Y) $ enters the see-saw
(type I) formula. It was shown\cite{ID} that knowledge of these combinations allows,  
in principle, one to reconstruct parameters of the  RH neutrino sector  (masses, phases).
For this, in turn, one needs to reconstruct completely
the mass matrix of light neutrinos, discover SUSY and
measure rare processes with high enough accuracy. This looks
practically impossible, at least now. 

Partial tests of the see-saw can be done by studying the 
rare decays induced by the slepton mixing. The branching ratio equals
\begin{equation}
B(\mu \rightarrow e \gamma) =
\frac{\alpha^3}{G_F^2 m_{SUSY}^8} |(m_S^2)_{\mu e}|^2 \tan^2\beta , 
\label{bratio}
\end{equation}
where $m_{SUSY} = m_{SUSY}(m_0, m_{1/2})$ is the effective SUSY mass parameter,
$m_{1/2}$ is the gaugino mass, $\tan \beta$ is the ratio of MSSM Higgs
doublet VEV's, and $(m_S^2)_{\mu e}$ is given in Eq.~(\ref{slepton}).

If the large lepton mixing originates from the Dirac mass matrix
(lopsided models, versions of the single RH neutrino dominance), the
Yukawa couplings $Y_{\mu i}$, $Y_{e i}$ are large and 
for $m_{SUSY} \sim 200$ GeV the branching ratio in Eq.~(\ref{bratio}) turns out to be
$10^{-12} - 10^{-11}$ -  at the level of the present experimental bound.
Determination of $m_{SUSY}$ in terms of
$m_0$ and  $m_{1/2}$  beyond the leading log approximation has further enhanced
the branching ratio\rlap{.}\,\cite{pet}

\subsection{\bf Other Mechanisms}

What are other possibilities  apart from the see-saw? The incomplete list includes. 

1. Various radiative mechanisms: The  Zee (one loop) mechanism\cite{zee} 
is essentially excluded in its minimal version by data\rlap{.}\,\cite{zeeH} 
One loop generation also occurs  in the SUSY models with 
trilinear  R-parity violating couplings. 
Neutrino masses can be generated in two loops as suggested by Zee\cite{zee} and  Babu\rlap{.}\,\cite{babu}
 
2. Neutrino mass generation  by the bi-linear R-parity violation terms\rlap{.}\,\cite{bili} 
This mechanism  is a combination of the see-saw and radiative effects:
neutrino mass appears as the see-saw due to mixing of neutrinos with
neutralinos (Higgsino) and the latter is generated by running from the high mass scales.

3. Mechanisms related to the existence of extra dimensions. There are different
scenarios: (i)  large extra dimensions (ADD)\cite{add} where the Dirac neutrino mass
is suppressed by  the large volume of extra dimensions,
(ii) warped extra dimensions (RS), where the RH neutrinos can be  zero 
modes localized on the hidden brane, thus leading again 
to the small Dirac neutrino mass\rlap{,}\,\cite{RS} and (iii)
infinite extra dimensions\rlap{.}\,\cite{DGP}

4). There are several new proposals which implement various 
realizations of the see-saw mechanism.

Models with dynamical symmetry breaking\cite{dyn}  reproduce the low scale 
see-saw mechanism with  the RH neutrino masses below the symmetry breaking scale,  and correspondingly, 
with small  Dirac masses (much smaller than the masses of quarks and charged leptons). 

Also  in models with ``Little Higgs'' \cite{little} the neutrinos get  masses  via 
the low scale see-saw. 

In models with dimensional deconstruction\cite{deconst}
the see-saw scale (masses of the RH neutrinos) is determined by the inverse lattice spacing.
The Dirac mass matrix is nearly diagonal and the large (maximal)
mixing follows from the pseudo-Dirac structures in the RH mass matrix
which correspond to $L_e - L_{\mu} - L_{\tau}$ symmetry.
Essentially, the maximal mixing appears because
different lepton  families belong to different sites of the lattice 
and the link scalar fields  (singlets of SM) couple with
the RH components of neutrinos, thus producing the off-diagonal
(in flavor space) Majorana mass terms.

These alternative mechanisms imply  deviation from 
minimality. They can accommodate the neutrino masses and produce some interesting features.
However, they do not really  lead  to a better understanding of the
experimental results and require introduction  of  additional elements (physics)
beyond  the Standard Model with RH neutrinos.

Turning the arguments: the neutrino data can be used to put limits on the
suggested alternative mechanisms, and consequently, on  physics beyond the SM.
In such a way the neutrinos can probe  extra dimensions, dynamical
symmetry breaking, etc..

\section{Conclusions}

During last several years enormous progress has been achieved in
the determination of the neutrino masses and mixings and in studies of the neutrino mass matrix.
Still, large freedom exists in the possible structures of the mass matrix  which
leads to very different interpretations of the  results. 
There are no definite hints from the bottom-up approach yet, and more information is needed, 
in particular, on the type of mass spectrum. 

The main question (still open) is: what is behind the obtained results?
What is the underlying physics? 
Preference? Probably, the  see-saw  associated to the Grand Unification. 
The context of $SO(10)$ looks rather appealing in spite of known problems.  
Other mechanisms (being in a less advanced stage of development) 
are not excluded and can give leading or sub-leading 
contributions to neutrino mass.

How can ideas about neutrinos be checked? Future experiments will perform
precision measurements of neutrino parameters. Apart from this
to understand the underlying physics we will
certainly need results from the non-neutrino experiments: 

- astrophysics and cosmology;

- searches for rare processes like 
flavor violating lepton decays,  proton decay, etc.; and 

- future high energy colliders.

\section*{Acknowledgments}

I am grateful to E. Kh. Akhmedov,  P. de Holanda, M. Frigerio, C. Lunardini  
and O. Peres  for fruitful discussions 
and help in preparation of this talk. 


\renewcommand{\theequation}{A.\arabic{equation}}



\end{document}